\documentclass[12pt,thmsa]{article}
\usepackage{amssymb}

\usepackage{sw20aip}

\input{tcilatex}
\begin{document}

\author{Gino Segr\`{e} \and Department of Physics, University of Pennsylvania \and %
Philadelphia, Pa. 19104 (email:segre@dept.physics.upenn.edu)}
\title{Cyclotron Emissivity in the Early Universe}
\date{March i8,1998 }
\maketitle

\begin{abstract}
Primordial magnetic fields present between the eras of neutrino decoupling
and of positron-electron annihilation can lead to the emission of
neutrino-anti neutrino pairs or axions by electrons in excited Landau
levels, thereby changing the electron energy spectrum. The rate is
calculated and shown to not affect significantly the standard Big Bang
picture.
\end{abstract}

In a neutron star, a large ambient magnetic field may cause electrons to
precess in cyclotron orbits. Under the proper conditions, electrons in
excited states corresponding to higher Landau levels decay to the ground
state by the emission of a neutrino-anti-neutrino pair which escapes from
the neutron star. This particular form of energy loss has been studied for
almost three decades, subject to refinements due to better models of neutron
stars and weak interactions .

The effects of a large primordial magnetic field on Big Bang Nucleosynthesis
(BBN) have also been considered for almost thirty years, starting with the
initial works of Matese and O'Connell\cite{matese}and of Greeenstein\cite
{greenstein}. This research centered on possible changes in the Universe's
expansion rate due to the magnetic field energy and on shifts in the weak
interaction rates due to changes in available phase space, degeneracy etc.%
\cite{cheng}\cite{rubinstein}\cite{starkman}.

As is well known, conditions inside neutron stars bear some resemblance to
those prevailing in the early Universe at the time of BBN\cite{raffelt}. It
therefore seems worthwhile to examine if the primordial magnetic fields
presumed extant in the early universe could have led to a degradation of
electron energy by cyclotron emissivity. It is clear that one needs to focus
on the period after neutrino decoupling since before then electrons and
positrons are in thermal equilibrium with neutrinos and anti-neutrinos and a
shift in the annihilation rate has no effect. It is equally clear that the
mechanism has negligible effect after the annihilation of electrons and
photons into photons since the electron(positron) population drops
dramatically at that point.\cite{kolb}

The period we focus on then corresponds to the roughly one hundred seconds
when the Universe drops from a temperature of $2\cdot 10^{10}K$ to $2\cdot
10^{9}K$, or equivalently from an energy of about three electron masses, $%
3m_{e},$ to an energy of one third of an electron mass,$\frac{m_{e}}{3}.$ 
\cite{kolb} In fact, during this time the primordial magnetic field is
decreasing as well: assuming the scale factor $R$ is related to temperature
by $RT$ remaining constant and assuming also that the magnetic flux remains
constant, we expect the magnetic field $B\backsim T^{2}$. $B$ changes by a
factor of one hundred as $T$ changes by a factor of ten.

The first studies of the effect of a primordial magnetic field on BBN
analyzed modifications of phase space due to the $B$ field and shifts in the
expansion rate of the Universe because of the added magnetic field energy.
These early calculations have been refined by several authors in
increasingly detailed calculations, not without some controversy. The
present limits, as quoted by Grasso and Rubinstein,\cite{rubinstein} are
that, for $T=10^{9}K,$ we must have $B\leq 10^{12}G$ (where of course we are
measuring $B$ in Gauss) for a magnetic field coherence length $L\ll
10^{11}cm.$ If the coherence length is taken to be the horizon length, the
limits on $B$ are more severe, i.e. $B\leqslant 10^{11}G.$ For our purposes,
we need the first limit since we only require that the magnetic field be
homogeneous over an electron cyclotron orbit. Assuming, as we said in the
previous paragraph, that $B\backsim T^{2}$, we see that at $T=10^{10}K$ ,
the maximum allowed magnetic field is $B=10^{14}G.$

The energy spectrum of an electron moving in a magnetic field directed along
the z axis is quantized according to the formula 
\begin{equation}
E(p_{z},n,s_{z})=\sqrt{p_{z}^{2}+m_{e}^{2}+eB(2n+1+s_{z})}
\end{equation}
where $n=0,1,2,3....$ label the energy levels, $p_{z}$ is the z component of
momentum and $s_{z}$ is the z component of the electron spin. Clearly the
larger the B field, the larger the energy released in a transition from a
higher to a lower energy level. On the other hand, if the B field is too
large, the higher levels are not excited thermally; the optimal value is $%
\frac{eB}{m_{e}T}\approx 1$ when $T\leq $ $m_{e}$ and $\frac{eB}{T^{2}}%
\approx 1$ otherwise.

Conventionally $B_{c}=4.4$ $\cdot 10^{13}G.$ is called the critical field .
It corresponds to a magnetic field $eB_{c}=m_{e}^{2}$ , clearly the order of
magnitude we are considering. We have already seen that our limits on $B$
indicate that $B\leqslant 2B_{c}$ for $T\approx 10^{10}K\approx 2m_{e}$ so
in fact we are in the regime where $\frac{eB}{m_{e}T}\approx 1$. We may
therefore expect cyclotron emissivity to play a significant role, but how
significant is the question. For convenient comparison with the literature,
we will quote temperature in units of $T_{9}$ which means multiples of $%
10^{9}$ $K$ and magnetic fields in units of $B_{13}$ , obviously multiples
of $10^{13}G.$ We normalize our energies to $\rho _{e}$ , the energy in the
electron field, which we take as\cite{kolb} 
\begin{equation}
\rho _{e}\approx 7\cdot 10^{22}T_{9}^{4}\frac{ergs}{cm^{3}}
\end{equation}
Using the above formula as a rough guideline,we can estimate the electron
energy density in the region of interest to us. For e.g. $T\sim 10^{10}K$ ,
we see that $\rho _{e}\sim 10^{27}\frac{ergs}{cm^{3}}$ and that $%
n_{e}\approx $ $10^{32}/cm^{3}$ , an electron density similar to that near
the edge of a neutron star. As an estimate of the energy loss $\ $due to
emission by electrons of neutrino-anti-neutrino pairs, we therefore use the
formulas derived for neutron stars by Kaminker, Levenfish and Yakolev.\cite
{kaminker} They introduce parameters 
\begin{equation}
\xi =\frac{T_{p}}{T\;}\quad \quad T_{p}\approx 2\cdot
10^{9}B_{13}X^{2}K.\quad \quad X_{F}=\frac{p_{F}}{m_{e}}
\end{equation}
and go on to consider the limits $\xi \ll 1$ and $\xi \gg 1.$ In our case,
the limit $\xi \gg 1$ is applicable (extrapolating the $\xi \ll 1$ gives a
comparable value) . In the $\xi \gg 1$ case the formula for the rate of
energy loss is 
\begin{equation}
Q_{\nu }\approx 10^{15}B_{13}^{2}T_{9}^{5}\frac{ergs}{cm^{3}\cdot \sec }.
\end{equation}

The best way to estimate the magnitude of the effect is by comparing $Q$ to $%
\dot{\rho}=\frac{d\rho }{dt}$ , the rate of change in electron energy
density due to the expansion of the universe (in the above and from now on
we denote $\rho _{e}$ simply as $\rho \ $). Using time as defined by this
expansion rate\cite{kolb} 
\begin{equation}
t=\left( .3g_{\bullet }^{\frac{-1}{2}}\right) \frac{m_{Planck}}{T^{2}}\sim
\left( \frac{T}{Mev}\right) ^{-2}
\end{equation}
we find that 
\begin{equation}
\frac{d\rho }{dt}\sim \frac{d\rho }{dT}\cdot \frac{dT}{dt}\sim \frac{2\rho }{%
t}.
\end{equation}
Since $B\sim T^{2},$we see that $Q_{\nu }\sim T_{9}^{9},$ while, using the
previous expressions for $\rho $ and for $t$ , we see that $\dot{\rho}\sim
T_{9}^{6}.$ Their ratio is therefore a function of $T$ or equivalently, of
time. Scaling $B$ by setting $B=10^{12}G$ for $T=10^{9}K$ and using $%
1Mev.=1.2\cdot 10^{10}K$ to scale $t$ ,we see that $\frac{Q_{\nu }}{\dot{\rho%
}}$ may be written as 
\begin{equation}
\eta _{\upsilon }=\frac{Q_{\nu }}{\dot{\rho}}\sim \frac{10^{13}T_{9}^{9}}{%
10^{20}T_{9}^{6}}\sim 10^{-7}T_{9}^{3}.
\end{equation}
The ratio of rates of electron energy decrease due to emission of
neutrino-anti-neutrino pairs and expansion of the universe ranges from about 
$10^{-3}$ for $T\sim 2\cdot 10^{10}K.$ to about $10^{-6}$ at
electron-positron annihilation. At this level, the result does not affect
the standard BBN model. Even if neutrinos decoupled at a temperature a
factor of three higher and if the present formula for $Q_{\nu }$ could be
extrapolated to those higher values of $B$ , the result for $\eta _{\nu }$
would not significantly affect BBN.

A few caveats/questions are in order: 1)is the formula for $Q_{\nu }$ valid?
The answer is probably yes within the limits of the calculation. As Kaminker
et al. emphasize, the result is insensitive to the value of the electron
Fermi momentum and the extrapolation to larger values of $B$ also appears
valid 2) could $B$ be even larger than $10^{12}G$ at $T\simeq 10^{9}K?$ Of
course this limit is only valid over a coherence length much smaller than
the horizon, but it is conceiveable that even larger of $B$ are present over
distances small enough to affect $\rho _{\upsilon }$ . Turning back to the
formula for the electron energy levels, we see that even for $p_{z}^{2}\sim
T^{2}\gtrapprox m_{e}^{2}$ , the electrons are all in the ground state and
hence do not radiate unless $eB\lesssim T^{2}$ . This corresponds to $%
B\lesssim 10^{12}G$, so we see that a larger magnetic field would not lead
to more significant limits.

We would like to turn now to a related topic, the limits on electron-axion
couplings imposed by the cyclotron emission of axions%
\cite[see this text for references to axions]{raffelt}. The situation is
much like the case we have just discussed. The corresponding $Q_{a}$ for $%
e\rightarrow e+a$ was derived by Borisov and Grishina \cite{borisov} . 
\begin{equation}
Q_{a}=1.6\cdot 10^{40}g_{ae}^{2}X_{F}^{\frac{-2}{3}}T_{9}^{\frac{13}{3}%
}B_{13}^{\frac{2}{3}}\frac{ergs}{cm^{3}\cdot \sec }
\end{equation}
where $g_{ae}$ is the axion-electron coupling constant and $X_{F}$ was
defined earlier.This formula has been recently recently reanalyzed by
Kachelriess et al.\cite{kachelriess} who have shown that it holds reasonably
well even for values of $B$ which are substantially larger than $B_{c}$ . In
the above equation we insert the previously discussed limit for the magnetic
field, $B\leq 10^{12}G$ at $T=10^{9}K$ and assume $B\sim T^{2}$. Using the
standard formulas for fermionic energy density and number density, we obtain 
$n_{e}\simeq 1.5\cdot 10^{28}T_{9}^{3}$ $cm.^{-3}.$ With the relation $%
p_{F}=\left( 3\pi ^{2}n_{e}\right) ^{\frac{1}{3}},$ we find $p_{F}\sim
0.4\cdot T_{9}m_{e}$.

Putting this all together we obtain 
\begin{equation}
\eta _{a}=\frac{Q_{a}}{\dot{\rho}}\simeq \frac{10^{40}g_{ae}^{2}T_{9}^{5}}{%
10^{20}T_{9}^{6}}=10^{20}g_{ae}^{2}T_{9}^{-1}
\end{equation}
Demanding $\eta _{a}\lessapprox 1$ leads to a bound on the axion-electron
coupling $g_{ae}\lessapprox 10^{-10}$ , compareable to the neutron star
bound obtained by Kachelriess et al.\cite{kachelriess} As one sees in the
above equations this bound does not depend drastically on the value of the
magnetic field: it is only weakened by a factor of order two if $B\sim
10^{11}G.$ at $T=10^{9}K.$ On the other hand the bound is much weaker than
that obtained by studying red giant stars\cite{raffelt} or even, as
emphasized by Kachelriess et al.,\cite{kachelriess}using the same process of
cyclotron emissivity for magnetic white dwarf stars. Finally we would like
to mention that similar types of calculations have been carried out by
Kuznetzov and Mikheev.\cite{kuznetsov}

We would like to thank the Department of Energy for support under grant
D.O.E. 3071.

\end{document}